\documentclass[unnumsec,webpdf,modern,large]{oup-authoring-template}%

\graphicspath{{Fig/}}
\usepackage{comment}

\theoremstyle{thmstyleone}%
\theoremstyle{thmstyletwo}%
\theoremstyle{thmstylethree}%

\usepackage{draftwatermark}
\usepackage{float}
\SetWatermarkText{Draft}
\usepackage{booktabs}
\usepackage{acro}
\acsetup{make-links = false}

\DeclareAcronym{SciML}{
  short = SciML,
  long  = scientific machine learning
}

\DeclareAcronym{UQ}{
  short = UQ,
  long  = uncertainty quantification
}
\DeclareAcronym{CNN}{
  short = CNN,
  long  = convolutional Neural Network,
  short-plural-form = CNNs,
  long-plural-form  = convolutional Neural Networks
}
\DeclareAcronym{MRST}{
  short = MRST,
  long  = MATLAB Reservoir Simulation Toolbox
}

\DeclareAcronym{MCMC}{
  short = MCMC,
  long  = Markov chain Monte Carlo
}

\DeclareAcronym{MAP}{
  short = MAP,
  long  = maximum a posteriori
}

\DeclareAcronym{OTWD}{
  short = OTWD,
  long  = optimal transport Wasserstein distance
}

\begin{document}

\journaltitle{PNAS Nexus}
\DOI{DOI added during production}
\copyrightyear{YEAR}
\pubyear{YEAR}
\vol{XX}
\issue{x}
\access{Published: Date added during production}
\appnotes{Paper}

\firstpage{1}


\title[SciML for \emph{``FluidFlower"}]{Learning and Inferring Multiphase Flow Dynamics in Porous Media using Scientific Machine Learning: Application to the \emph{``FluidFlower"} CO$_2$ Injection Experiment}

\author[1,$\ast$]{Hannah Lu}
\author[2]{Llu\'{i}s Sal\'{o}-Salgado}
\author[1]{Yun-Ting Chou}
\author[3]{Ehsan Haghighat}
\author[3,4]{Ruben Juanes}

\address[1]{\orgdiv{Oden Institute for Computational Engineering and Sciences}, \orgname{The University of Texas at Austin}, \orgaddress{201 E 24th St, Austin, TX 78712, USA}}

\address[2]{\orgdiv{Department of Earth, Environmental, and Planetary Sciences}, \orgname{The University of Tennessee, Knoxville}, \orgaddress{1621 Cumberland Avenue, Knoxville, TN 37996, USA}}

\address[3]{\orgdiv{Department of Civil Environmental Engineering}, \orgname{Massachusetts Institute of Technology}, \orgaddress{77 Massachusetts Avenue, Cambridge, MA 02139, USA}}

\address[4]{\orgdiv{Department of Earth, Atmospheric, and Planetary Sciences}, \orgname{Massachusetts Institute of Technology}, \orgaddress{77 Massachusetts Avenue, Cambridge, MA 02139, USA}}

\corresp[$\ast$]{To whom correspondence should be addressed: \href{mailto:hannah.lu@austin.utexas.edu}{hannah.lu@austin.utexas.edu}}

\received{Date}{0}{Year}
\revised{Date}{0}{Year}
\accepted{Date}{0}{Year}


\abstract{Accurate prediction and parameter identification of multiphase flow in porous media are central challenges in geological carbon dioxide storage, due to strong nonlinearities, high-dimensional parameter spaces, and limited observational data. We present a scientific machine learning
framework that integrates surrogate modeling and Bayesian inference to enable efficient forward prediction and inverse parameter estimation for CO$_2$-brine flows in geological media. The approach is demonstrated in the \emph{``FluidFlower"} experimental rig, a controlled laboratory system that provides high-resolution, time-resolved observations of CO$_2$ migration in heterogeneous porous media. A convolutional neural network surrogate is trained on high-fidelity numerical simulations to learn the evolution of CO$_2$ saturation and dissolved CO$_2$ concentration fields for a wide range of values characterizing the multiphase flow properties of different geologic layers. The trained surrogate is then embedded within a Markov chain Monte Carlo framework to perform parameter inference conditioned on experimental observations. Our results show that the surrogate accurately captures large-scale CO$_2$ plume migration, CO$_2$ dissolution dynamics, and multiphase flow behavior while providing six orders-of-magnitude acceleration compared to traditional simulations. Embedding the surrogate within a Bayesian inference framework enables computationally tractable exploration of the high-dimensional parameter space and reveals both identifiable and non-identifiable parameter combinations that produce similar macroscopic plume behavior. By leveraging spatially and temporally resolved full-field experimental observations, the proposed framework substantially improves the macroscopic concordance between simulations and experiments compared to the previous manually calibrated results based on limited plume-scale quantities. Analysis using progressively increasing observation horizons further demonstrates that observational data become substantially more informative once the plume interacts with major geological features such as faults and sealing layers.} 


\keywords[Abbreviations]{GCS, SciML, MRST, CNN, MCMC, UQ, MAP, OTWD}

\otherabstract[Significance Statement]{Accurate prediction of subsurface multiphase flow is essential for applications such as geological carbon storage, but remains difficult because of complex flow physics and uncertain geological properties. This work shows how scientific machine learning can bridge observational data and numerical simulations by enabling rapid prediction and making uncertainty-aware inference computationally feasible for multiphase flow systems. The results demonstrate the value of assimilating spatially and temporally resolved observations to substantially improve model predictive accuracy. These advances provide an important step toward practical digital twins for improving prediction and supporting decision-making in large-scale subsurface energy and environmental systems.
}

\maketitle


\section{Introduction}
Geological carbon dioxide storage (GCS) has emerged as a key strategy for mitigating anthropogenic emissions, supported by decades of research~\cite{benson2005geological,benson2008co2,bachu2000sequestration, krevor2023subsurface} and growing field experience~\cite{sharma2011co2crc,preston2005iea,hovorka2006measuring,luth2020geophysical,niemi2020characterizing}. When injected into deep geological formations such as saline aquifers, CO$_2$ migrates through porous rock under the combined influence of pressure gradients, buoyancy, and subsurface heterogeneity, leading to complex plume dynamics that are well studied~\cite{juanes2006impact,hesse2008gravity,macminn2011co2}. A range of trapping mechanisms contribute to long-term storage, including structural trapping beneath caprocks, residual trapping within pore spaces, dissolution into brine, and mineralization over longer timescales~\cite{benson2005geological,benson2008co2}.  At larger spatial scales, the effectiveness of these mechanisms depends on plume migration and its interaction with formation properties, which control the spatial distribution and extent of trapping~\cite{szulczewski2012lifetime}. Building on this understanding, recent studies have demonstrated that substantial storage capacity exists globally, especially in offshore sedimentary basins with favorable geological conditions, highlighting the potential for large-scale deployment of GCS as part of climate mitigation strategies~\cite{ringrose2019maturing}.

Despite decades of advances in multiphase flow modeling, reliable prediction and calibration of subsurface CO$_2$ migration remain  challenging. Process-based models, typically derived from conservation laws governing multiphase flow and transport, have become increasingly sophisticated, incorporating multiscale heterogeneity and coupled physical processes~\cite{aziz1979petroleum,pruess1999tough2,benson2008co2}. However, such high-fidelity models are computationally expensive and often require fine spatial and temporal resolution, making them impractical for uncertainty quantification, inverse modeling, or real-time decision-making tasks that demand repeated simulations. Moreover, their predictive reliability is difficult to assess due to the scarcity of field-scale observational data. Subsurface processes occur over large spatial and temporal scales, while available measurements are sparse, indirect, and often limited to well locations. This mismatch creates significant challenges in model validation and calibration, and ultimately in communicating model predictions and associated uncertainties to stakeholders~\cite{scheer2021subsurface}. As a result, there remains a gap between high-fidelity numerical simulations and actionable insights for risk assessment and decision-making in GCS.

In this context, controlled laboratory experiments provide a critical bridge between theory, simulation, and field-scale applications. The \emph{``FluidFlower"} experiment~\cite{ferno2024room} has emerged as a unique intermediate-scale benchmark~\cite{nordbotten2022final} for studying multiphase flow in porous media under well-characterized conditions. By combining high-resolution spatial measurements with reproducible experimental protocols, \emph{``FluidFlower"} enables systematic comparison between observations and numerical simulations, offering a rare opportunity to evaluate model fidelity and quantify discrepancies. The associated international benchmark study has revealed that, even under controlled conditions, significant mismatches can arise between simulations and experimental observations, reflecting uncertainties in key parameters such as permeability and capillary pressure as well as limitations in representing fine-scale flow dynamics~\cite{flemisch2024fluidflower}. While history matching can substantially improve agreement by calibrating model parameters, this process is inherently ill-posed and often requires iterative, computationally intensive simulations guided by expert judgment~\cite{haugen2023physical,salo2024direct,tian2024history}. Consequently, despite improved post-calibration performance, the cost and non-uniqueness of history matching pose significant challenges for scalable prediction and real-time decision support.

These challenges motivate the development of \ac{SciML} approaches that can complement or partially replace traditional simulation workflows. In particular, surrogate models trained on high-fidelity simulations offer a promising route to approximate complex multiphase flow dynamics at a fraction of the computational cost, enabling rapid forward prediction across high-dimensional parameter spaces~\cite{wen2023real,mo2019deep,zhong2019predicting,lu2025uncertainty,lu2025lithological}. At the same time, recent advances in machine learning for inverse problems have introduced scalable approaches for parameter inference, including surrogate-assisted Bayesian methods and deep generative models that enable efficient exploration of posterior distributions~\cite{tang2020deep, mosser2018stochastic,siahkoohi2023reliable,jagalur2018inferring}. Despite these developments, most existing studies often rely on synthetic datasets without systematic validation against physical experimental observations. As a result, the extent to which SciML-based approaches can simultaneously support accurate forward modeling and reliable parameter inference in realistic subsurface systems remains an open question.

In this work, we develop a \ac{SciML}-based framework for forward surrogate modeling and inverse parameter inference of multiphase flow in porous media, with application to the \emph{``FluidFlower"} CO$_2$ injection experiments. Our approach leverages a \ac{SciML} surrogate model trained on high-fidelity simulations to approximate the evolution of CO$_2$ saturation and dissolved CO$_2$ concentration fields under uncertain subsurface parameters. This surrogate is then embedded within a Bayesian inference framework to enable efficient inverse identification of key parameters using experimental observations. Leveraging the unique availability of spatially resolved, time-dependent data from the \emph{``FluidFlower"} experiments, we systematically evaluate both the predictive accuracy of the surrogate model and its effectiveness in supporting parameter inference. The results demonstrate that the surrogate accurately captures large-scale plume dynamics while enabling six orders-of-magnitude acceleration. Data generation and training takes some efforts which can be automated in the backend, but the trained model empowers engineers to perform simulation, \ac{UQ}, and inverse analysis near real time. At the same time, the inferred parameter distributions reveal both identifiable and non-identifiable directions in the parameter space, providing insight into the limitations of calibration under realistic data constraints. We further investigate how the observational time horizon impacts the predictive performance of the calibrated model, providing insight into when plume observations become most informative for model calibration. These capabilities provide a key building block toward the development of data-driven digital twins~\cite{willcox2023foundational} for subsurface systems, enabling efficient integration of models and observations for improved prediction and decision support.

\section{Results}
We evaluate the proposed \ac{SciML} framework for surrogate modeling and parameter inference using the medium-size \emph{``FluidFlower"} rig~\cite{flemisch2024fluidflower,salo2024direct}. The analysis focuses on three key aspects of the surrogate model: (i)~its accuracy in learning forward multiphase flow dynamics, (ii)~its ability to infer uncertain parameters from observational data, and (iii)~its predictive capability under uncertainty.

\subsection{Forward Surrogate Modeling of Multiphase Flow}

We first evaluate the ability of the \ac{SciML} surrogate model to learn multiphase flow dynamics from high-fidelity simulations (the forward problem). The surrogate predicts the evolution of both CO$_2$ saturation and dissolved CO$_2$ concentration fields directly from sampled geological parameter realizations, enabling efficient full-field reconstruction of the system dynamics.

\begin{figure*}[!hbp]%
\centering
\includegraphics[width = \textwidth]{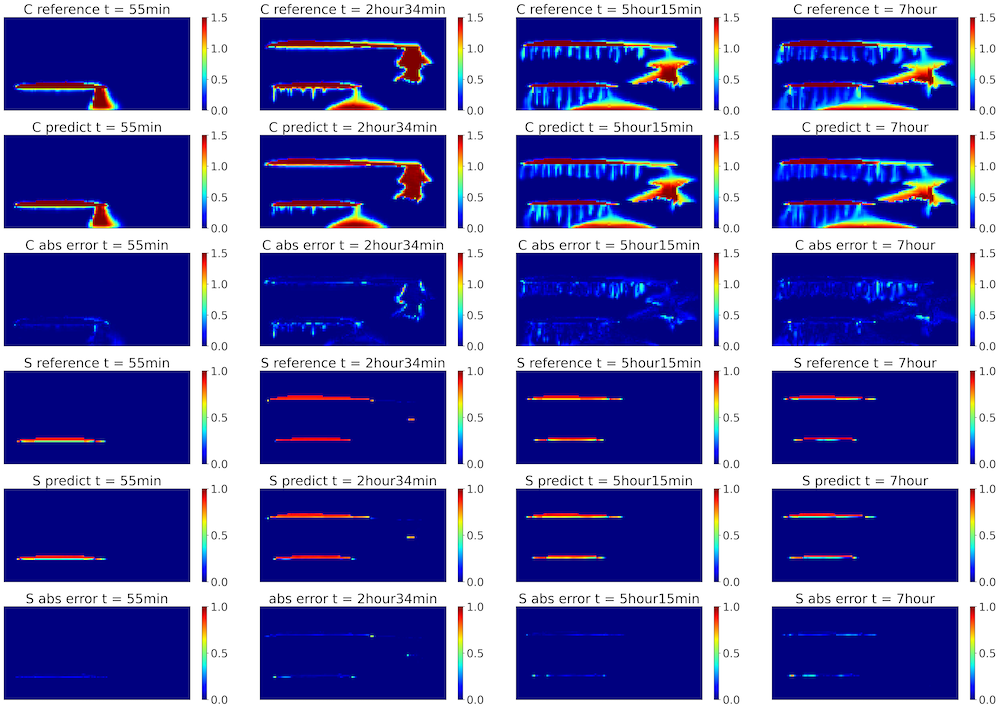}
\caption{Surrogate model predictions of forward multiphase flow dynamics for an unseen test case at t = 55 min, 2 h 34 min, 5 h 15 min, and 7 h. Top three rows: dissolved CO$_2$ concentration fields in kg$/$m$^2$. Bottom three rows: CO$_2$ saturation fields. In each group, the first row shows the reference MRST simulation, the second row shows the SciML surrogate prediction, and the third row shows the absolute error.}\label{fig:test1}
\end{figure*}

Figure~\ref{fig:test1} presents representative predictions for an unseen test case with parameter combinations not included in the training dataset. The figure compares reference simulations, surrogate predictions, and corresponding pointwise absolute errors at multiple time instances for both dissolved CO$_2$ concentration and CO$_2$ saturation fields. Overall, the surrogate accurately captures the large-scale multiphase flow behavior, including buoyancy-driven upward migration, lateral spreading, and the interaction of the plume with heterogeneous geological structures, while providing more than six orders-of-magnitude acceleration compared to the original high-fidelity simulations.

For the dissolved CO$_2$ concentration fields, the model successfully reproduces the overall organization of dissolution patterns and transport dynamics over time. In particular, the surrogate captures preferential migration pathways and accumulation behavior induced by fault and seal structures. For the CO$_2$ saturation fields, the surrogate accurately predicts the sharp gas plume interface, plume extent, and temporal evolution of the mobile CO$_2$ phase. The predicted saturation distributions remain in strong agreement with the reference simulations across all reported times, demonstrating the ability of the model to learn coupled multiphase transport behavior.

At finer spatial scales, discrepancies are observed in the detailed fingering structures in the dissolved CO$_2$ concentration fields, as highlighted by the pointwise error maps. These differences are expected because dissolution fingering is governed by hydrodynamic instabilities that are highly sensitive to small perturbations and local nonlinear dynamics, making the precise spatial location of individual fingers inherently difficult to predict. Consequently, pointwise errors tend to concentrate in regions with active fingering dynamics and near sharp moving interfaces.

\begin{figure*}[!htp]%
\centering
\includegraphics[width = \textwidth]{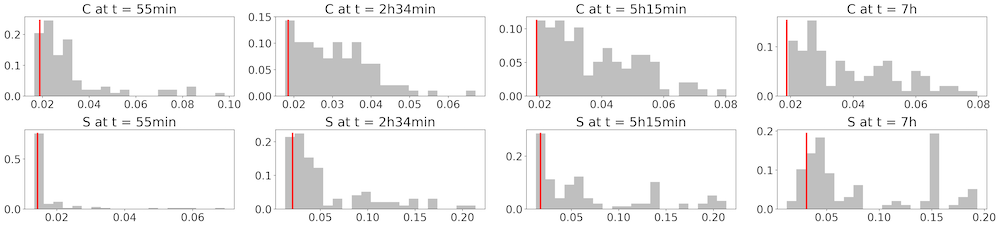}
\caption{Quantitative error analysis for dissolved CO$_2$ concentration (top row) and CO$_2$ saturation (bottom row) at four representative times. Gray histograms show the distribution of OTWD between the test case and all training samples, while the red vertical line indicates the OTWD between the surrogate prediction and the reference simulation for the test case.}\label{fig:err1}
\end{figure*}

To quantitatively evaluate predictive performance, we compute the two-dimensional \ac{OTWD} between surrogate predictions and reference simulations for both dissolved CO$_2$ concentration and CO$_2$ saturation fields. Because \ac{OTWD} measures discrepancies between spatial distributions rather than pointwise pixel differences, it provides a physically meaningful metric for transport-dominated multiphase flow systems involving moving interfaces and fingering dynamics. Figure~\ref{fig:err1} compares the \ac{OTWD} between the surrogate prediction and the reference simulation against the distribution of \ac{OTWD} between the test case and all training realizations at representative times. The gray histograms characterize the intrinsic variability and diversity of the dataset by measuring the similarity between the unseen test case and the training simulations. At early times, the distributions remain relatively concentrated, indicating that the plume structures across different realizations are still broadly similar. At later times, however, the distributions become substantially wider and increasingly complex due to the emergence of nonlinear fingering dynamics and heterogeneous multiphase flow patterns. This evolution indicates that the test case cannot be interpreted as a trivial interpolation of nearby training samples, particularly during the later stages of plume evolution when the system dynamics become increasingly complex. The red vertical line denotes the \ac{OTWD} between the surrogate prediction and the reference simulation for the selected test case. For both dissolved CO$_2$ concentration and CO$_2$ saturation, the surrogate prediction errors remain consistently near the lower end of the histogram distributions across all reported times, even as the diversity of the dataset increases significantly at later times. These results indicate that the surrogate predictions remain substantially closer to the reference solution than a typical training realization.

We further evaluate predictive accuracy relative to the intrinsic variability of the dataset using a relative Wasserstein error metric $\epsilon_\text{rel}$, defined as the ratio between the surrogate prediction error and the average \ac{OTWD} between the test realization and all training samples (see Eq.~\ref{eq:rel-err} in Materials and Methods). Values of $\epsilon_\text{rel}<1$ indicate that the surrogate prediction error is smaller than the average Wasserstein distance between the test realization and the training realizations. Table~\ref{tab:rw} shows that the relative Wasserstein error $\epsilon_\text{rel}$ generally decreases over time. Although the variability among realizations increases substantially at later times due to the emergence of nonlinear fingering and heterogeneous plume dynamics, the surrogate maintains robust predictive performance relative to the increasing variability and complexity of the solution space.

\begin{table}[!h]
\centering
\caption{Relative Wasserstein error $\epsilon_\text{rel}$ at representative times for the test case.}
\label{tab:rw}
\begin{tabular}{l|cccc}
\toprule
& 55 min &2 h 34 min &5 h 15 min &7 h \\
\midrule
concentration & 0.6002 &0.6043 &0.5003  &0.4778\\
saturation & 0.7959 &0.3762 &0.2472 &0.3674\\
\bottomrule
\end{tabular}
\end{table}

Despite localized discrepancies in fine-scale fingering structures, the surrogate predictions remain highly accurate at the scale relevant for system-level analysis, capturing the dominant transport behavior, plume evolution, and multiphase flow structure. Both the qualitative comparisons and Wasserstein-distance-based quantitative analyses demonstrate that the \ac{SciML} surrogate maintains strong predictive performance even in increasingly complex nonlinear flow regimes. This level of fidelity is sufficient for downstream tasks such as inverse parameter inference and uncertainty-aware prediction, where the primary objective is to accurately capture macroscopic system behavior rather than exact fine-scale instability patterns. Additional test cases exhibiting similarly strong predictive performance, including cases with substantially different plume migration and fingering behavior, are provided in Figure~\ref{fig:test2} of the Supplementary Information.

\subsection{Inverse Parameter Inference from Observations}

Compared with the international benchmark study~\cite{flemisch2024fluidflower}, where no history matching was performed, the calibrated simulations in~\cite{salo2024direct,landa2026performance} demonstrate significantly improved concordance between numerical predictions and experimental observations. Here, \emph{concordance} refers to the agreement between simulated and observed system behavior~\cite{oldenburg2018we}. The identified parameter sets enable reduction of uncertainty in petrophysical properties and provide insight into the governing physical processes.

We next evaluate the ability of the proposed framework to infer uncertain subsurface parameters from experimental observations. The inverse analysis is performed using Bayesian inference with the learned \ac{SciML} surrogate as a computationally efficient forward model. This enables rapid exploration of the high-dimensional parameter space while avoiding the prohibitive computational cost associated with repeated high-fidelity simulations.

\begin{figure*}[!hbp]%
\centering
\includegraphics[width = \textwidth]{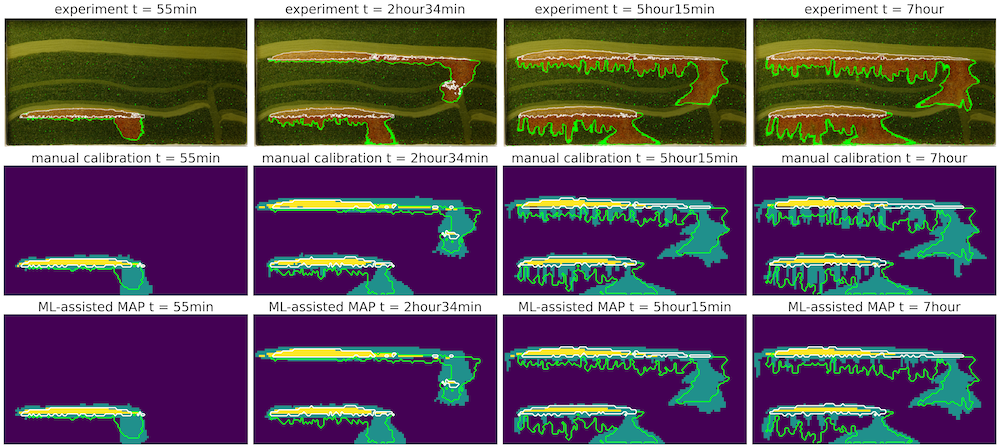}
\caption{Comparison of experimental observations, manual calibration results, and ML-assisted calibration at 55 min, 2 h 34 min, 5 h 15 min, and 7 h. Top row: experimental images with extracted dissolved CO$_2$ (green) and gaseous CO$_2$ (white) plume contours. Middle row: MRST simulations evaluated at the manually calibrated parameter set. Bottom row: MRST simulations evaluated at the inferred MAP parameters obtained from surrogate-assisted MCMC inversion.}\label{fig:calibration}
\end{figure*}

\begin{figure*}[!hbp]
\centering
\includegraphics[width = \textwidth]{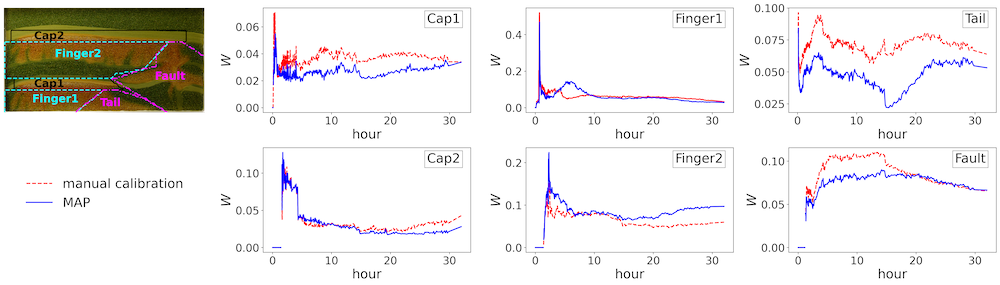}
\caption{Evolution of the Wasserstein distance between the experimental observations and simulated plume distributions in six representative regions of the domain (left panel). Results from the manually calibrated parameter set (red dashed line) are compared with those from the ML-assisted MAP estimate (blue solid line). }\label{fig:regional_error}
\end{figure*}

Figure~\ref{fig:calibration} compares experimental observations with predictions obtained from the manually calibrated parameter set~\cite{salo2024direct} and the inferred \ac{MAP} parameter set identified through surrogate-assisted \ac{MCMC} inversion. In~\cite{salo2024direct}, the calibration was performed manually through a few repeated forward simulations, where model parameters were iteratively adjusted to match a limited set of representative observations and global plume-scale quantities, including plume areas at selected times and characteristic fingering arrival times. Specifically, the matching procedure relied on quantities such as the areal extent of gaseous and dissolved CO$_2$ within selected reservoir regions at $t=55$min and $t=154$min, together with the times at which dissolution fingers reached specific locations in the domain.   Because each full-physics simulation required substantial computational cost, the calibration did not utilize the complete spatiotemporal evolution of the full-field experimental observations. In contrast, our surrogate-assisted Bayesian inversion framework enables inference directly from the full time series of full-field experimental data, making it possible to enhance the concordance of the spatial morphology and temporal evolution of the plume simultaneously. The top row shows the processed experimental observations, including extracted dissolved CO$_2$ and gaseous CO$_2$ plume boundaries. The middle and bottom rows show corresponding high-fidelity simulations evaluated at the manually calibrated in~\cite{salo2024direct} and our surrogate inferred MAP parameter sets, respectively. 

\begin{figure*}[!hbp]
\centering
\includegraphics[width = \textwidth]{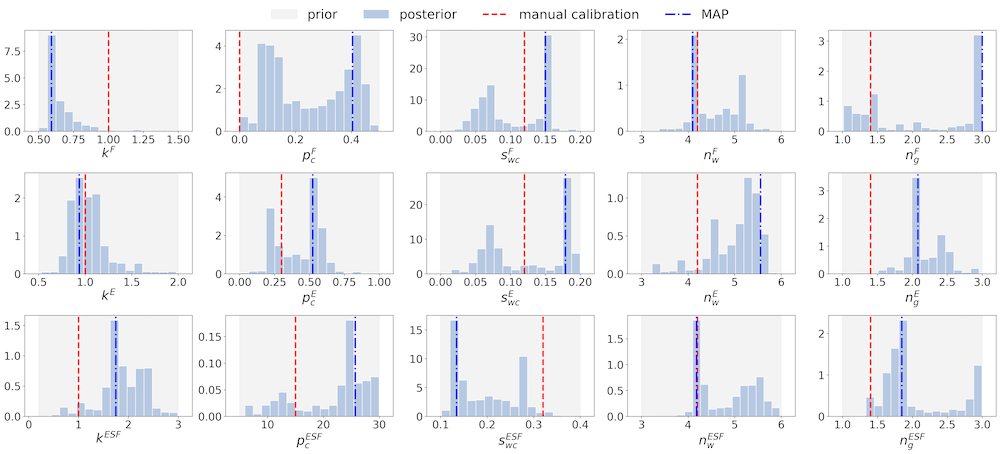}
\caption{Posterior distributions obtained from MCMC sampling with random initialization. Subplots are arranged consistently with Table~\ref{tab:params}, with rows representing geological units and columns representing parameter categories. Gray shaded regions indicate prior support, blue histograms denote posterior samples, red dashed lines indicate manually calibrated parameter values, and blue dash-dotted lines indicate MAP estimates.}\label{fig:posterior}
\end{figure*}

Overall, the inferred MAP solution reproduces several important macroscopic features of the observed plume evolution and demonstrates improved concordance in plume morphology relative to the manually calibrated solution. The improvement is most evident in the large-scale plume geometry and migration pathways. In particular, the inferred MAP solutions better capture the smaller vertical extent of dissolved CO$_2$ at early time ($t = 55$min), the reduced vertical plume accumulation near the fault region, and the overall plume geometry observed in the experiments at intermediate and later times. These improvements are reflected quantitatively in Figure~\ref{fig:regional_error}, where the MAP solution exhibits substantially lower Wasserstein distances in the Tail and Fault regions over much of the observation period. The inferred solution also more accurately reproduces the lateral spreading behavior and plume shapes across multiple geological layers, indicating that the full-field inversion better constrains the interaction between buoyancy-driven migration and heterogeneous geological structures. While the manually calibrated solution reproduces some fine-scale dissolution finger patterns more closely, this may partially reflect the calibration strategy adopted in~\cite{salo2024direct}, where fingering arrival times at selected locations were included among the calibration targets. This trend is also reflected in Figure~\ref{fig:regional_error}, where the manually calibrated solution generally exhibits lower Wasserstein distances in the Finger1 and Finger2 regions, suggesting that the calibration strategy of~\cite{salo2024direct} is particularly effective at matching fingering-related plume features. Remaining discrepancies of both solutions are primarily associated with fine-scale fingering dynamics and localized interface variations, which are highly sensitive to small perturbations and difficult to constrain uniquely from observation data. In addition, \cite{salo2024direct} pointed out that some discrepancies in dissolution fingering may arise from limitations of the underlying physics-based simulator itself, including the absence of mechanical dispersion, which can influence the width, morphology, and migration speed of convective fingers.  Since the surrogate model is trained to emulate the simulator, these model-form deficiencies are inherited by both the surrogate and the subsequent inversion, limiting the extent to which calibration can improve agreement in such fine-scale features.

Figure~\ref{fig:posterior} shows the marginal posterior distributions obtained from \ac{MCMC} sampling with random initialization. The posterior distributions exhibit varying degrees of concentration across parameters, indicating that some parameters are more strongly constrained by the observations than others. Several parameters (e.g., permeability multiplier $k$ in fault region E, connate water saturation $s_\text{wc}$ in storage region F, and water relative permeability exponent $n_w$ in top seal region ESF) exhibit relatively concentrated posteriors centered near the manually calibrated values, suggesting that they can be identified from the available observations. In contrast, other parameters remain broadly distributed or strongly skewed, indicating weaker constraints and substantial posterior uncertainty. For example, the posterior distribution of the permeability multiplier $k$ in storage region F shows strong updates relative to the prior ranges, reflecting the sensitivity of plume migration to this property. The inferred posterior for the gas relative permeability exponent $n_g$ in storage region F remains partially identifiable, with the manually calibrated value located near the lower tail of the posterior distribution. Several posterior modes deviate substantially from the manually calibrated parameter values while still producing comparable plume evolution and spatial patterns. This behavior reflects the inherently ill-posed nature of history matching in this system, where distinct parameter combinations can generate similar macroscopic observations. In particular, different combinations of permeability, capillary pressure, and relative permeability parameters can compensate for one another and produce similar large-scale plume migration dynamics.

\subsection{Effect of Observation Horizon on Predictive Performance}
To investigate how the observation horizon influences predictive performance, we perform a series of calibrations using progressively increasing portions of the experimental time series. Specifically, the model is calibrated using observations up to four different times: $t = 55$min, $t = 2$h$34$min, $t = 5$h$15$min, and $t = 7$h. For each observation horizon, the inferred parameters are subsequently used to predict the plume evolution over the entire simulation horizon.

\begin{figure*}[!hbtp]%
\centering
\includegraphics[width = \textwidth]{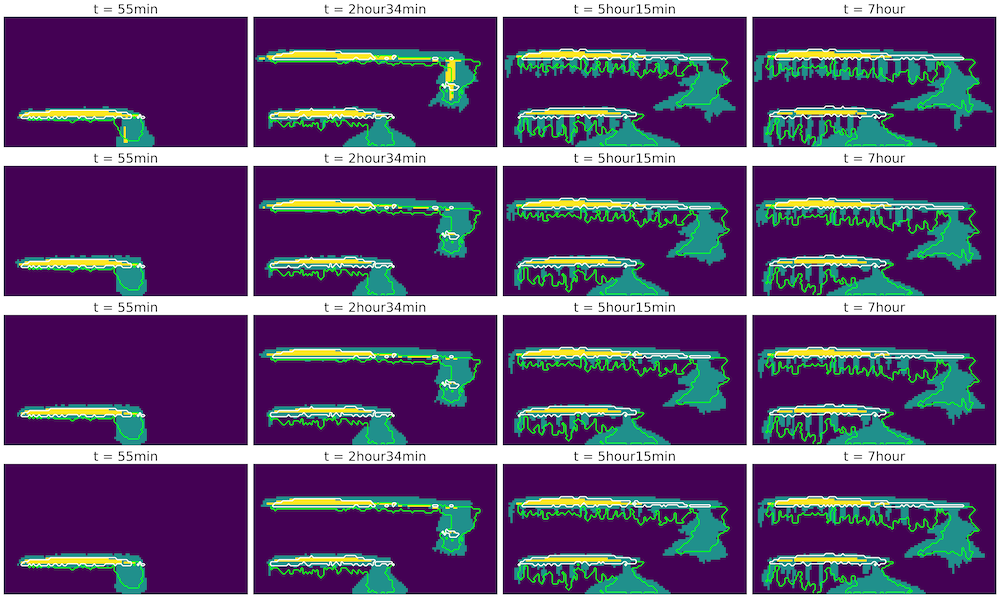}
\caption{Model calibration results using progressively increasing observation horizons. Rows correspond to parameter estimates inferred using experimental observations up to $t = 55$ min, $t = 2$ h $34$ min, $t = 5$ h $15$ min, and $t = 7$ h, respectively. Columns show the predicted plume states at representative times. Green and white contours denote the experimentally extracted dissolved and gaseous CO$_2$ plume boundaries.}
\label{fig:calibration_vs_t}
\end{figure*}

Figure~\ref{fig:calibration_vs_t} compares the resulting plume predictions at representative times. Each row corresponds to a different observation horizon used for calibration, while the columns show the predicted plume states at $t = 55$ min, $t = 2$ h $34$ min, $t = 5$ h $15$ min, and $t = 7$ h, respectively. The green and white contours denote the experimentally extracted dissolved and gaseous CO$_2$ plume boundaries.

\begin{figure*}[!htbp]
\centering
\includegraphics[width = \textwidth]{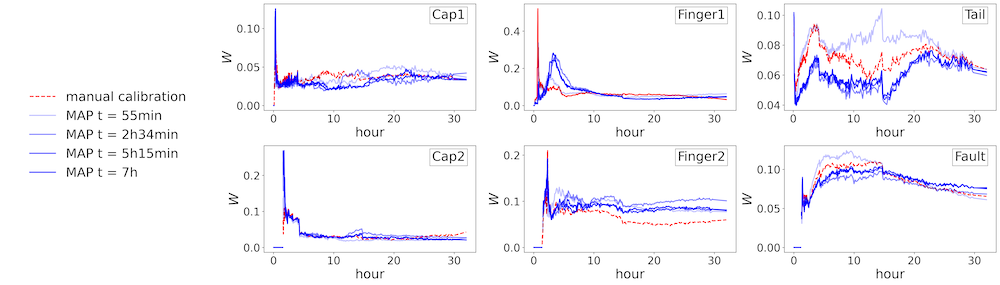}
\caption{Evolution of the Wasserstein distance between the experimental observations and simulated plume distributions in six representative regions of the domain (left panel of Figure~\ref{fig:regional_error}). Results from the manually calibrated parameter set (red dashed line) are compared with those from the ML-assisted MAP estimates inferred using experimental observations up to  $t = 55$ min, $t = 2$ h $34$ min, $t = 5$ h $15$ min, and $t = 7$ h, respectively (blue solid line). }\label{fig:regional_error_vs_t}
\end{figure*}

When calibration is performed using only the early-time observations ($t \leq 55$min), the inferred model captures the near-injection ($I_1$ in Figure~\ref{fig:tank}) plume behavior reasonably well but exhibits noticeable discrepancies at later times, particularly in the large-scale migration pathways and downward dissolution structures induced by $I_2$ (see Figure~\ref{fig:tank}). At this stage, the observations mainly contain information associated with the first CO$_2$ injection in the lower region of the domain, before the plume has substantially interacted with the upper geological layers and fault structures. Consequently, the calibration primarily constrains parameters associated with the lower storage region, while parameters representing the sealing and fault materials remain comparatively weakly identified. This trend is also reflected quantitatively in Figure~\ref{fig:regional_error_vs_t}, where the Wasserstein distances decrease substantially when the observation horizon is extended from $t=55$ min to $t=2$ h $34$ min and after, particularly in the Tail and Fault regions that are directly influenced by the later plume migration pathways and interactions with geological heterogeneities.

Extending the observation horizon to $t = 2$ h $34$min substantially improves the predicted plume evolution. By this time, the experimental observations contain information associated with the second injection in the upper region, and the plume has reached and interacted with the fault structure. These additional dynamics provide significantly stronger constraints on the permeability and capillary properties associated with the fault and sealing materials. As a result, the inferred model more accurately reproduces the macroscopic plume geometry, lateral spreading behavior, and interaction with the layered geological structures. The improvement is particularly evident in the large-scale dissolved CO$_2$ migration patterns and the extent of gaseous plume accumulation beneath the sealing layers.

Interestingly, further increasing the observation horizon beyond $t = 2$ h $34$ min yields comparatively modest improvements in predictive performance, despite producing somewhat different inferred parameter combinations. Consistent with this observation, Figure~\ref{fig:regional_error_vs_t} shows only modest additional reductions in Wasserstein distance when the observation horizon is further extended to t=5 h 15 min and t=7 h. The predicted plume morphology remains largely consistent across the later calibration windows, suggesting that the dominant transport behavior has already been sufficiently constrained once the plume establishes its primary migration pathways and interacts with the major geological heterogeneities. The remaining discrepancies at later times are primarily associated with fine-scale fingering structures and instability-driven plume features, which are highly sensitive to small perturbations and difficult to reproduce exactly even under similar parameter settings, and, as discussed previously, may also reflect limitations of the underlying physics-based simulator rather than parameter uncertainty alone~\cite{salo2024direct}. Consequently, while additional late-time observations continue to influence the inferred parameters, they do not lead to substantially improved large-scale predictive behavior.

\section{Discussion}

This work demonstrates that scientific machine learning can provide an effective bridge between high-fidelity multiphase flow simulation and physical observations for uncertainty-aware prediction and parameter inference in subsurface systems. A particularly important observation is that the surrogate maintains strong predictive performance even as the solution space becomes increasingly complex at later times due to nonlinear fingering dynamics and heterogeneous plume migration. The Wasserstein-distance analysis in Figure~2 provides additional insight into this behavior. At early times, the \ac{OTWD} distributions between different realizations remain relatively concentrated because the plume evolution is still dominated by large-scale buoyancy-driven transport and the flow structures across parameter realizations remain broadly similar. At later times, however, the distributions become substantially wider and increasingly complex due to the emergence of instability-driven fingering dynamics and heterogeneous interactions with geological structures. This behavior highlights an important distinction between \ac{SciML} learned multiphase flow dynamics and simple interpolation between training samples. Conventional interpolation-based approaches may perform reasonably well when different realizations remain structurally similar, but their accuracy would be expected to deteriorate as the underlying solution manifold becomes increasingly nonlinear and diverse. In contrast, the \ac{SciML} surrogate predictions remain consistently closer to the reference solution than a typical training realization even in the later-time regime characterized by complex plume evolution and fingering growth. These results suggest that the surrogate is learning physically meaningful transport behavior and parameter-dependent flow dynamics rather than merely reproducing nearby samples from the training dataset. At the same time, the remaining discrepancies remain concentrated in highly localized fingering structures and sharp moving interfaces, reflecting the intrinsic sensitivity of these instability-driven features to small perturbations and local nonlinear dynamics.

The results also highlight several important limitations and directions for future improvement. The current surrogate architecture does not explicitly exploit temporal correlations in the evolving plume dynamics and relies on a relatively limited training dataset due to the substantial computational cost and frequent nonlinear convergence failures of the full-physics simulations~\cite{flemisch2024fluidflower,salo2024direct}. In addition, the original MRST simulations defined on unstructured PEBI grids were interpolated onto a uniform $64 \times 128$ Cartesian grid for compatibility with CNN architectures, potentially losing some fine-scale geometric information associated with local refinement near faults and injection regions. Future work could therefore benefit from improved temporal modeling strategies, larger and more systematically sampled training datasets, and mesh-aware learning approaches such as graph neural networks~\cite{pfaff2020learning,ju2024learning} or operator-learning methods~\cite{kovachki2023neural,wen2022u} that directly operate on unstructured simulation grids.

The inverse modeling results additionally illustrate the fundamentally ill-posed nature of history matching in subsurface multiphase flow systems. Multiple parameter combinations can generate similar macroscopic plume behavior, particularly when observations are limited, noisy, indirect, or spatially coarse. The existence of multiple plausible solutions underscores the importance of exploring posterior distributions rather than relying solely on a single calibrated parameter set. In this context, the surrogate model plays a particularly important role by enabling efficient exploration of the high-dimensional parameter space, which would be computationally infeasible using repeated full-physics \ac{MRST} simulations alone. Direct Bayesian exploration of the 15-dimensional parameter space with the original simulator would require an enormous number of expensive forward evaluations, further complicated by frequent nonlinear convergence failures. By replacing the full simulator with a computationally efficient surrogate, the proposed framework makes large-scale posterior exploration practically feasible while preserving the dominant multiphase flow dynamics. This capability opens the door to further investigation of physically plausible parameter combinations and uncertainty structures, which can subsequently be analyzed in greater detail using targeted high-fidelity simulations, comparison with experimental observations, and domain-specific physical interpretation.

The sensitivity of the posterior distributions to different random initializations of the \ac{MCMC} ensemble further highlights the nonuniqueness of the inverse problem. As shown in the supplementary Figures~\ref{fig:calibration1}-\ref{fig:calibration2}, different initializations can lead to somewhat different inferred parameter combinations while still producing calibrated plume predictions with substantially better concordance with the experimental observations than the manual calibration. Despite this variability, several parameters exhibit relatively consistent behavior across different initializations. In particular, the permeability multiplier $k^{E}$ remains comparatively well identified and stays close to the manually calibrated reference value across all runs, suggesting that this parameter is strongly constrained by the observed large-scale plume migration behavior. In contrast, the connate water saturation $s_\text{wc}^{ESF}$ consistently exhibits a substantial shift away from the manually calibrated value, indicating that the experimental observations strongly favor a different mobility behavior in the top seal region. Several parameters associated with the ESF sealing material, including $k^{ESF}$ and $p_c^{ESF}$, also show systematic posterior updates toward larger values, reflecting the important role of the upper sealing layers in controlling plume spreading and trapping behavior once the plume reaches the upper region of the domain. At the same time, other parameters remain broadly distributed or multimodal across different initializations, suggesting weaker identifiability and stronger parameter compensation effects. These observations emphasize that multiple parameter combinations can achieve similar levels of macroscopic concordance with the experimental observations, particularly when the remaining discrepancies are dominated by instability-driven fingering structures and localized plume dynamics.

The analysis using progressively increasing observation horizons also provides insight into the evolving information content of observations during plume migration. In particular, the results suggest that the most informative stage for parameter identification occurs once the plume begins interacting with the dominant geological heterogeneities, including the upper sealing layers and fault structures. After this stage, additional observations appear to provide diminishing returns for constraining the dominant large-scale transport behavior. Instead, the remaining differences between realizations become increasingly associated with instability-driven fingering dynamics and localized plume structures that are intrinsically sensitive to small perturbations and difficult to reproduce uniquely. This behavior suggests that the information content of observations may partially saturate once the primary migration pathways and macroscopic plume organization become established, with important implications for monitoring design, online calibration, and real-time forecasting in subsurface flow systems. In particular, the results indicate that simply incorporating increasingly large amounts of late-time full-field data may not necessarily lead to substantially improved large-scale predictive capability, especially when the remaining discrepancies are dominated by highly sensitive fingering dynamics. This observation raises important questions regarding the optimal selection and representation of observational data for history matching and uncertainty quantification. Rather than relying exclusively on dense full-field observations, it may be more effective to combine processed spatial information with carefully selected global quantities that capture physically meaningful plume behavior, such as plume extent, migration rates, dissolution patterns, or trapping characteristics. Indeed, the better agreement of the manually calibrated solution in the Finger1 and Finger2 regions of Figure~\ref{fig:regional_error} and Figure~\ref{fig:regional_error_vs_t} confirms that carefully selected physic-informed summary quantities may provide more effective constraints than dense full-field observations alone. Developing principled strategies for identifying the most informative observations and integrating multi-scale data representations remains an important direction for future work, particularly for practical monitoring and adaptive calibration workflows in large-scale subsurface energy systems.

Overall, the proposed surrogate-assisted Bayesian framework demonstrates the potential of combining full-physics simulation, experimental observation, and \ac{SciML} for the development of uncertainty-aware digital twins in subsurface flow systems. A particularly important aspect of the present study is the use of spatially and temporally resolved full-field experimental observations, which provide substantially richer information than the sparse measurements typically available in reservoir-scale applications. The availability of full-field plume evolution data enables simultaneous calibration of plume morphology, migration pathways, and large-scale transport dynamics, allowing the inverse analysis to constrain physically meaningful subsurface parameters that would be difficult to identify from limited scalar observations alone. At the same time, practical reservoir- and basin-scale systems rarely provide such dense observational coverage, raising important questions regarding how these approaches can be extended to realistic field settings. Future work will therefore investigate the impact of sparse and partially observed data by systematically masking or subsampling the available full-field observations to emulate realistic monitoring configurations. Such studies may help identify which subsets of observations contain the most informative content for parameter identification and forecasting, while also providing guidance for optimal monitoring design and adaptive data acquisition strategies. More broadly, the present results provide an important step toward data-driven digital twins for subsurface energy and environmental systems~\cite{willcox2023foundational,gahlot2024digital}. Beyond uncertainty-aware forecasting and parameter inference, future developments may incorporate feedback control, adaptive monitoring, and sequential data assimilation to enable real-time decision support during subsurface operations. Another important direction involves investigating the transferability and generalization capability of the learned surrogate models across different geological stratigraphies, experimental tanks, and injection scenarios. Understanding how \ac{SciML} models trained in one setting can be adapted or transferred to related subsurface systems will be critical for scaling digital twin methodologies from controlled laboratory experiments toward realistic field-scale applications.

\section{Materials and Methods}

\subsection{Problem setup and data}
\begin{figure}[!htp]%
\centering
\includegraphics[width = 0.45\textwidth]{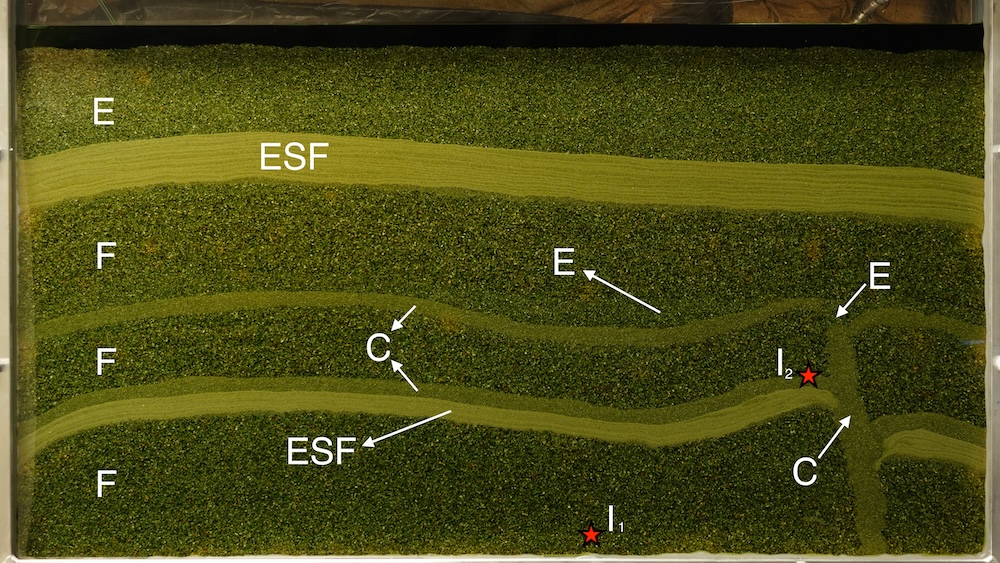}
\caption{Front view of porous medium in Tank 1, with lithologies in white and injector location shown with a red star. }
\label{fig:tank}
\end{figure}

\paragraph{Experimental Setup and Observational Data} 
A distinct element of our work is the use of physical experiments of CO$_2$ injection conducted using the \emph{``FluidFlower"} rigs \cite{ferno2024room}. These rigs are meter-scale, quasi-2D tanks with transparent Plexiglass panels designed and built in-house at the University of Bergen that replicate, by using unconsolidated sands with different grain sizes, geologic settings that are conducive to storing CO$_2$ underground, including complex structures such as folds and faults (Figure~\ref{fig:tank}). Here, we employ data from experiments carried out in Tank~1 (dimensions: $90\times47\times1.1$~cm). The rigs have multiple ports that allow the fluids to be flushed out after a given CO$_2$ injection, such that multiple injections can be conducted in the same setting. Because the \emph{``FluidFlower"} rigs are transparent quasi-2D tanks, they allow direct observation of two fields distributed over the entire domain: (1)~the CO$_2$ saturation (volume fraction of the pore space occupied by CO$_2$ gas), and (2)~the CO$_2$ concentration dissolved in water, a quantitative metric enabled by the presence of a pH indicator \cite{nordbotten2024darsia}. In this study, a total of $N_\text{obs} = 358$ image snapshots acquired over a period of 32 hours 4 minutes are used for model calibration.

\paragraph{Full-physics Modeling and Simulation} The computational data for training the surrogate models are generated by simulating ``full-physics'' models of multiphase flow in porous media that capture injection, migration, trapping and dissolution of CO$_2$ in a high-fidelity representation of the geologic structure. The flow physics are described by the mass balance equations for two-phase, two-component flow, supplemented with the multiphase extension of Darcy's law, with hysteretic capillary pressure and relative permeability functions \cite{bear2013dynamics} to capture residual trapping of CO$_2$ \cite{juanes2006impact} and partial miscibility between the fluids to capture the convective dissolution of CO$_2$ in water \cite{szulczewski2012lifetime} (see Figure~\ref{fig:test1}). This set of nonlinear partial differential equations is solved for pressure, saturation, and concentration using finite-volume spatial discretization and finite-difference time stepping \cite{lie2019introduction, schlumberger2014, salo2024three}. The resulting nonlinear algebraic system is solved at each time step using Newton’s method, with each iteration requiring the solution of a large linear system. The convergence of Newton's method can be challenging, especially if (as is the case in the \emph{``FluidFlower"} experiments) the fluid properties such as density and viscosity exhibit large contrasts~\cite{flemisch2024fluidflower, salo2024direct}.

\paragraph{Training Data Description} The numerical model for the \emph{``FluidFlower"} has been well set up in our recent work~\cite{salo2024direct} using the \ac{MRST}~\cite{krogstad2015mrst,lie2019introduction,lie2021advanced}. It is ready to be used to generate  data for our surrogate training tasks. Based on our direct comparison between \ac{MRST} numerical simulations and the \emph{``FluidFlower"} experimental data in~\cite{salo2024direct}, we identified $N_{\boldsymbol \gamma} = 15$ key parameters in the governing equation that play a critical role in history matching of the simulation prediction and experimental observation. The parameters $\boldsymbol \gamma\in \Gamma\subset \mathbb R^{N_\gamma}$ consist of 5 flow properties ($k$: permeability multiplier; $p_c$: capillary entry pressure in mbar; $s_\text{wc}$: connate water saturation; $n_w$: water relative permeability exponent; $n_g$: gas relative permeability exponent) in 3 material regions (F: storage sand; E: sand in fault; ESF: top seal). We assume a uniform prior distribution $\pi_\text{pr}$ for $\boldsymbol\gamma$, representing our best knowledge of the petrophysical characteristics, with ranges (see Table~\ref{tab:params}) informed by published measurements in similar silica sands (e.g.,~\cite{beard1973influence,smits2010thermal}) and constitutive laws (e.g., Kozeny-Carman~\cite{beard1973influence,trevisan2014investigation} and Brooks-Corey~\cite{brooks1965hydraulic} models). Let $\{\boldsymbol \gamma^{(i)}\}_{i=1}^{M}$ denote the $M$ samples from $\pi_\text{pr}$, we generate simulation outputs $\{c(\boldsymbol \gamma^{(i)},t_n), s(\boldsymbol \gamma^{(i)},t_n)\}_{i,n = 1}^{M,N}$ from MRST. Here $c$ is dissolved CO$_2$ concentration field and $s$ is gas CO$_2$ saturation field and $t_n$'s are non-uniform snapshot times, with the final experiment time given by $t_N = 48$ hours.
\begin{table}[!h]
\centering
\caption{Prior ranges of uncertain parameters $\boldsymbol{\gamma} \in \Gamma \subset \mathbb{R}^{15}$.}
\label{tab:params}
\begin{tabular}{lccccc}
\toprule
Parameter & $k$ & $p_c$  & $s_{\mathrm{wc}}$&$n_w$&$n_g$  \\
\midrule
F            & $[0.5,1.5]$ & $[0,0.5]$ & $[0,0.2]$ & $[3,6]$&$[1,3]$\\
E      & $[0.5,2]$ & $[0,1]$ & $[0,0.2]$&$[3,6]$&$[1,3]$ \\
ESF & $[0.2,3]$ & $[5,30]$ & $[0.1,0.4]$&$[3,6]$&$[1,3]$ \\
\bottomrule
\end{tabular}
\end{table}

\paragraph{Challenges in Data Generation} The simulator requires very small time-steps (on the order of seconds to minutes) due to the buoyancy of CO$_2$ at atmospheric conditions and high sand permeabilities.  The convergence of the nonlinear solver is very challenging---often requiring many iterations and frequent time-step cuts, which significantly increases computational cost.  Depending on the parameter combinations, each completed simulation can take anywhere from 3 hours to 3 days. A substantial number of parameter combinations result in incomplete runs due to convergence failures. These challenges can be summarized in two key aspects: (1) dense sampling in 15-dimensional space is computationally infeasible due to the curse of dimensionality---for instance, using just 5 sample points per dimension results in approximately $5^{15} \approx 3$ billion combinations; and (2) the difficulty in achieving nonlinear convergence leads to both a high failure rate and extremely long runtimes for some simulations. Together, these issues severely limit the ability to perform global sensitivity analysis and systematic parameter identification.

\subsection{Surrogate model for forward prediction}
The goal is to learn a mapping from $(\boldsymbol{\gamma}, t)$ to the dissolved CO$_2$ concentration field $c(\boldsymbol{\gamma}, t)$ and gas CO$_2$ saturation field $s(\boldsymbol{\gamma}, t)$ using the dataset of input-output pairs $\{(\boldsymbol{\gamma}^{(i)}, t_n) \mapsto c(\boldsymbol{\gamma}^{(i)}, t_n)\}^{i=1,\ldots,M}_{n=1,\ldots,N}$. The learned model is expected to generalize to unseen parameter configurations $\boldsymbol{\gamma} \in \Gamma$ and arbitrary times $t \in [0, t_N]$. Due to the computational challenges described above, we only have $M = 100$ completed simulation runs. We split the dataset into $M_{\text{train}} = 98$ runs for training and $M_{\text{test}} = 2$ runs for testing, one of which corresponds to the manually calibrated reference trajectory.  With $N = 466$ collected snapshots per run, this yields $98\times 466 = 45{,}668$ input-output pairs for training.

\begin{figure*}[!hbp]%
\centering
\includegraphics[width = \textwidth, trim=0 0 0 400, clip]{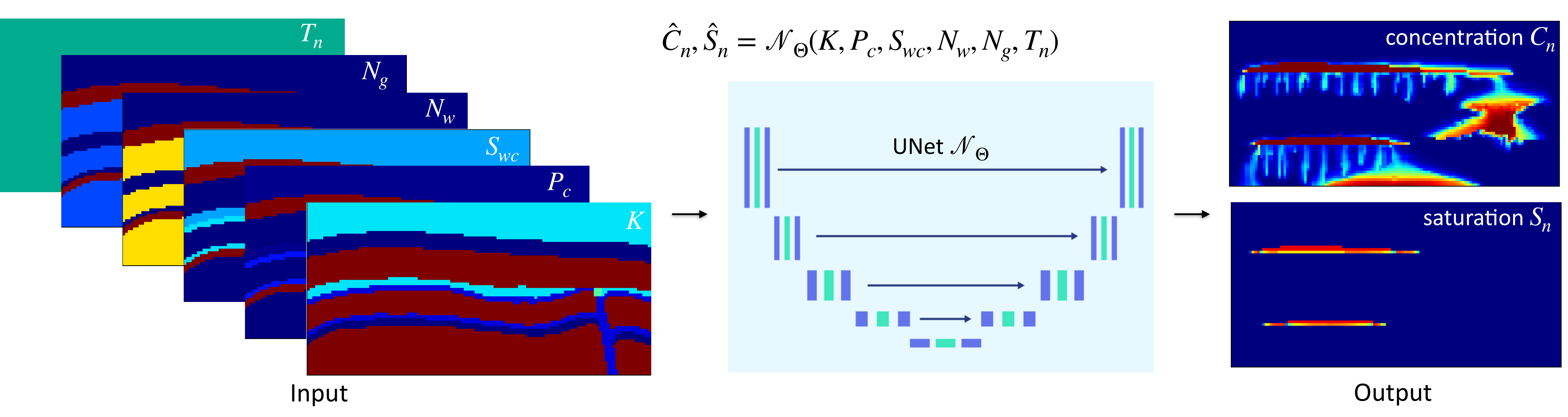}
\caption{Schematic illustration of the surrogate modeling framework. Inputs: spatially structured parameter fields together with time; Outputs: corresponding dissolved CO$_2$ concentration and CO$_2$ saturation fields.}
\label{fig:input-output}
\end{figure*}

To leverage spatial correlations, we endow the $15$-dimensional parameter vector $\boldsymbol{\gamma}$ with spatial structure from the \emph{``FluidFlower"} tank, yielding 5 input images ($K$, $P_c$, $S_\text{wc}$, $N_w$ and $N_g$) corresponding to the underlying material properties. For computational convenience and compatibility with \ac{CNN}, all input and output fields are represented on a uniform $64 \times 128$ Cartesian grid. The original simulation data, defined on an unstructured grid with perpendicular-bisector (PEBI) refinement (7038 cells, with local refinement near faults and wells), are interpolated onto this grid using nearest-neighbor interpolation. Treating time $t_n$ as an additional input channel (denoted by $T_n$), the learning task is formulated as an image-to-image regression problem (Figure~\ref{fig:input-output}).

We employ a \ac{CNN} architecture, UNet, tailored
for image-to-image regression~\cite{ronneberger2015u}. The network is trained to minimize the discrepancy between predicted and simulated concentration and saturation fields using a mean squared error (MSE) loss:
\begin{equation}
\mathcal{L} = \frac{1}{NM_\text{train}} \sum_{i=1}^{M_\text{train}}\sum_{n=1}^{N} \| \hat{C}_n^{(i)} - C_n^{(i)} \|_2^2 + \| \hat{S}_n^{(i)} - S_n^{(i)} \|_2^2,
\end{equation}
where $C_n^{(i)}$ and $\hat{C}_n^{(i)}$ denote the reference and predicted dissolved CO$_2$ concentration fields for the $i$-th training sample at $t_n$, and $S_n^{(i)}$ and $\hat{S}_n^{(i)}$ denote the corresponding reference and predicted CO$_2$ saturation fields at $t_n$.

\subsection{Evaluation metrics}

To further quantify predictive performance, we evaluate the surrogate using the two-dimensional \ac{OTWD} between predicted and reference fields for both dissolved CO$_2$ concentration and CO$_2$ saturation. Unlike pointwise metrics such as mean squared error, the \ac{OTWD} compares spatial distributions by measuring the minimal transport cost required to map one field onto another. As a result, it is substantially less sensitive to small spatial displacements of sharp interfaces or fingering structures, which can otherwise produce large pointwise errors despite preserving the correct large-scale transport behavior. This property is particularly important for multiphase flow systems, where hydrodynamic instabilities and nonlinear plume dynamics can lead to localized spatial shifts that are physically similar but pixelwise misaligned. Consequently, a small \ac{OTWD} indicates that the surrogate accurately captures the overall spatial organization, plume morphology, and transport dynamics of the system, even when fine-scale structures are not reproduced at exactly the same spatial locations.

Given two normalized spatial distributions $\mu$ and $\nu$ defined on the computational grid, the two-dimensional Wasserstein distance is computed as
\begin{equation}
W(\mu,\nu) = \min_{\Gamma \in \Pi(\mu,\nu)}
\sum_{i,j} \Gamma_{ij} d_{ij},
\end{equation}
where $\Pi(\mu,\nu)$ denotes the set of admissible transport plans satisfying the marginal constraints associated with $\mu$ and $\nu$, $\Gamma_{ij}$ represents the amount of mass transported from grid location $i$ to location $j$, and $d_{ij}$ denotes the Euclidean distance between the corresponding spatial coordinates. In this work, the concentration and saturation fields are first flattened into discrete probability distributions after normalization, and the ground distance matrix $d_{ij}$ is constructed using the Euclidean distances between pixel coordinates on the $64 \times 128$ grid. The Wasserstein distances are computed using the Python Optimal Transport (POT) library~\cite{flamary2021pot}, which provides efficient numerical solvers for discrete optimal transport problems.

To evaluate predictive accuracy relative to the intrinsic variability of the dataset, we additionally define a relative Wasserstein error metric. Let $W_{\mathrm{pred}}$ denote the OTWD between the surrogate prediction and the corresponding reference simulation for a test realization, and let
\begin{equation}
\bar{W}_{\mathrm{train}}
=\frac{1}{M_{\mathrm{train}}}
\sum_{i=1}^{M_{\mathrm{train}}}
W(U_{\mathrm{test}},U_i)
\end{equation}
denote the average OTWD between the test realization and all training realizations, where $U$ represents either the dissolved CO$_2$ concentration field or the CO$_2$ saturation field. The relative Wasserstein error is then defined as

\begin{equation}\label{eq:rel-err}
\epsilon_\text{rel}
=
\frac{W_{\mathrm{pred}}}{\bar{W}_{\mathrm{train}}}.
\end{equation}
Smaller values of $\epsilon_\text{rel}$ indicate that the surrogate prediction error is smaller than the average Wasserstein distance between the test realization and the training realizations.

In this work, the concentration and saturation fields are first flattened into discrete probability distributions after normalization, and the ground distance matrix $d_{ij}$ is constructed using the Euclidean distances between pixel coordinates on the $64 \times 128$ grid. The Wasserstein distances are computed using the Python Optimal Transport (POT) library~\cite{flamary2021pot}, which provides efficient numerical solvers for discrete optimal transport problems.

\subsection{Bayesian inference for parameter identification}

To address the challenges associated with high-dimensional parameter uncertainty and computationally expensive full-physics simulations, we adopt a Bayesian inference framework for uncertainty-aware parameter identification. Let $c_{\mathrm{exp}}(t)$ and $s_{\mathrm{exp}}(t)$ denote the experimentally observed dissolved CO$_2$ concentration and gaseous CO$_2$ saturation fields, respectively. The posterior distribution of the uncertain parameter vector $\boldsymbol{\gamma}$ is given by
\begin{equation}
\pi_{\mathrm{pos}}(\boldsymbol{\gamma} \mid c_{\mathrm{exp}}, s_{\mathrm{exp}})
\propto
\pi(c_{\mathrm{exp}}, s_{\mathrm{exp}} \mid \boldsymbol{\gamma})
\,
\pi_{\mathrm{pr}}(\boldsymbol{\gamma}),
\end{equation}
where $\pi_{\mathrm{pr}}(\boldsymbol{\gamma})$ denotes the prior distribution specified in Table~\ref{tab:params} and $\pi(c_{\mathrm{exp}}, s_{\mathrm{exp}} \mid \boldsymbol{\gamma})$ is the likelihood function.

The experimental observations consist of time-resolved binary maps indicating the spatial regions occupied by dissolved and gaseous CO$_2$. These binary maps are obtained through image processing of the experimental photographs by segmenting the characteristic color patterns associated with dissolved and gaseous CO$_2$ regions (see the top row of Figure~\ref{fig:calibration}). The surrogate model provides continuous predictions for dissolved CO$_2$ concentration $\hat{C}$ and CO$_2$ saturation $\hat{S}$, which are converted into binary maps using thresholding:
\begin{equation}
\begin{aligned}
&\hat{C}_{\mathrm{bin}}
=
\mathbf{1}
\left(
\hat{C}
\geq
C_{\mathrm{thr}}
\right),\\
&\hat{S}_{\mathrm{bin}}
=
\mathbf{1}
\left(
\hat{S}
\geq
S_{\mathrm{thr}}
\right),
\end{aligned}
\end{equation}
where $C_{\mathrm{thr}} = 0.21$ and $S_{\mathrm{thr}} = 0.001$ are the reference thresholds used to identify dissolved and gaseous plume regions, respectively~\cite{salo2024direct}.

The likelihood function is constructed based on the mismatch between the experimental binary maps and the thresholded surrogate predictions over all observation times. Assuming independent Gaussian errors, the likelihood takes the form
\begin{equation}
\begin{aligned}
\pi(c_{\mathrm{exp}}, s_{\mathrm{exp}} \mid \boldsymbol{\gamma})
\propto
&\exp
\left[
-\frac{1}{2\sigma^2}
\sum_{n=1}^{N_\text{obs}}
\left(
\left\|
c_{\mathrm{exp}}(t_n)
-
\hat{c}_{\mathrm{bin}}(\boldsymbol{\gamma}, t_n)
\right\|_2^2 \right.\right.\\
&+
\left.\left.\left\|
s_{\mathrm{exp}}(t_n)
-
\hat{s}_{\mathrm{bin}}(\boldsymbol{\gamma}, t_n)
\right\|_2^2
\right)
\right],
\end{aligned}
\end{equation}
where $\sigma$ controls the tolerance to mismatch between the observations and surrogate predictions. In the present numerical tests, we set $\sigma = 0.5$.

The posterior distribution is characterized using \ac{MCMC} sampling with the affine-invariant ensemble sampler implemented in the Python package \texttt{emcee}~\cite{foreman2013emcee}. The ensemble sampler evolves multiple interacting walkers simultaneously, which improves sampling efficiency and robustness in high-dimensional parameter spaces with correlated parameters. In the present implementation, we use $32$ walkers, slightly more than twice the parameter dimension, and run the chains for $1000$ steps while evaluating the surrogate model at each iteration. To assess the sensitivity of the inferred posterior to initialization and the potential existence of multiple plausible modes, the walkers are initialized randomly within the prior ranges. The resulting samples are used to estimate marginal posterior distributions shown in Figure~\ref{fig:posterior}, providing insight into the identifiability of individual parameters.

As a reference point, we additionally compute a \ac{MAP} estimate,
\begin{equation}
\boldsymbol{\gamma}_{\mathrm{MAP}}
=
\arg\max_{\boldsymbol{\gamma}}
\pi_{\mathrm{pos}}
(\boldsymbol{\gamma} \mid c_{\mathrm{exp}}, s_{\mathrm{exp}}),
\end{equation}
which provides a single best-fit parameter set for comparison with the posterior distributions and experimental observations.


\section{Competing interest}
The authors declare that they have no competing interests.

\section{Funding}
This work was funded by ExxonMobil through the ExxonMobil–MIT collaborative project ``Modeling and Mitigation of Induced Seismicity and Fault Leakage during CO2 storage". This work was supported in part by Department of Energy under award DE-SC0023171 (H. Lu) and the Air Force Office of Scientific Research under award FA9550-26-1-0002 (H. Lu).

\section{Author contributions}

H.L. conceived the methodology, developed the surrogate modeling and Bayesian inference framework, performed the surrogate training and inverse parameter identification, analyzed the results, and wrote the manuscript. L.S.-S. provided the experimental data, developed the MRST simulation setup and manual calibration workflow, contributed to the interpretation of the results, and contributed to writing the manuscript. E.H. and Y-T.C. contributed to code development and computational support. R.J. contributed to conceptual development, interpretation of the results, scientific discussion, and writing and reviewing the manuscript.

\section{Data availability}
The data and code underlying this study will be made publicly available in a GitHub repository (\url{https://github.com/Lu-s-Lab-UT-Austin/SciML-FluidFlower}) upon publication.

\section{Ethics statement}
This study does not involve human participants, animals, or personal data. The work is based on laboratory-scale multiphase flow experiments and numerical simulations associated with the \emph{``FluidFlower"}~\cite{salo2024direct}.

\bibliographystyle{unsrt}
\bibliography{main}

\clearpage
\begin{appendices}
\section{Supplementary Information}
\begin{figure*}[!hbp]%
\centering
\includegraphics[width = \textwidth]{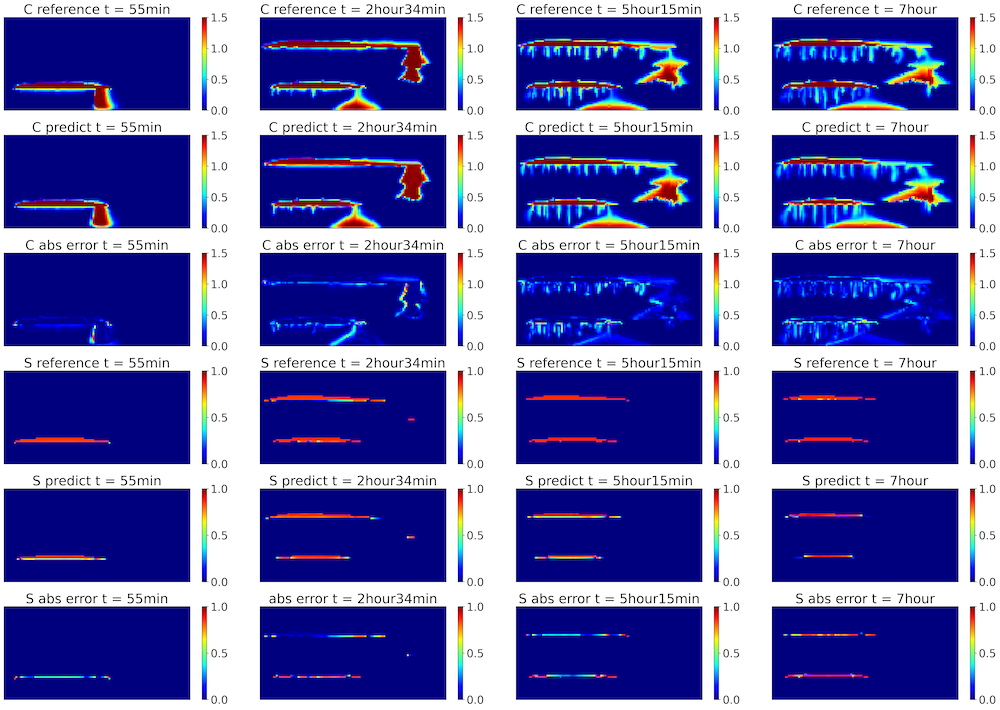}
\caption{Surrogate model predictions of forward multiphase flow dynamics for another unseen test case (different than Figure~\ref{fig:test1}) at t = 55 min, 2 h 34 min, 5 h 15 min, and 7 h. Top three rows: dissolved CO$_2$ concentration fields. Bottom three rows: CO$_2$ saturation fields. In each group, the first row shows the reference MRST simulation, the second row shows the SciML surrogate prediction, and the third row shows the absolute error.}\label{fig:test2}
\end{figure*}

\begin{figure*}[!htbp]%
\centering
\includegraphics[width = \textwidth, trim=0 0 0 555, clip]{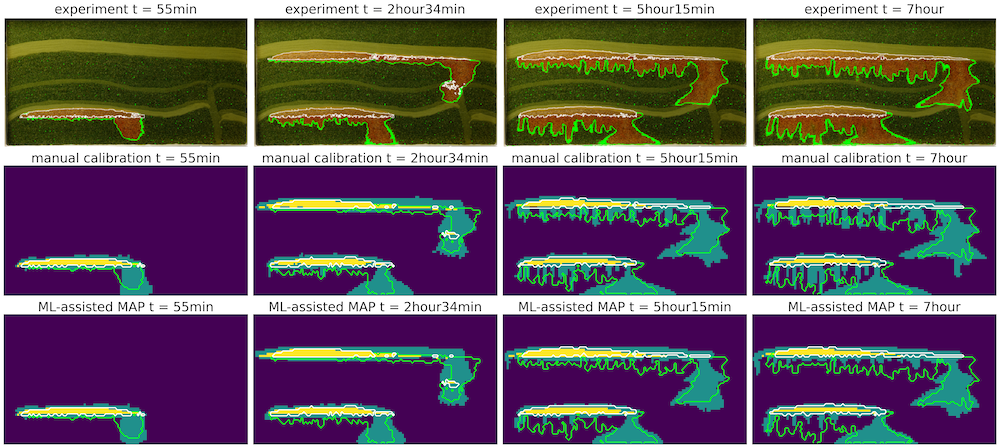}
\includegraphics[width = \textwidth]{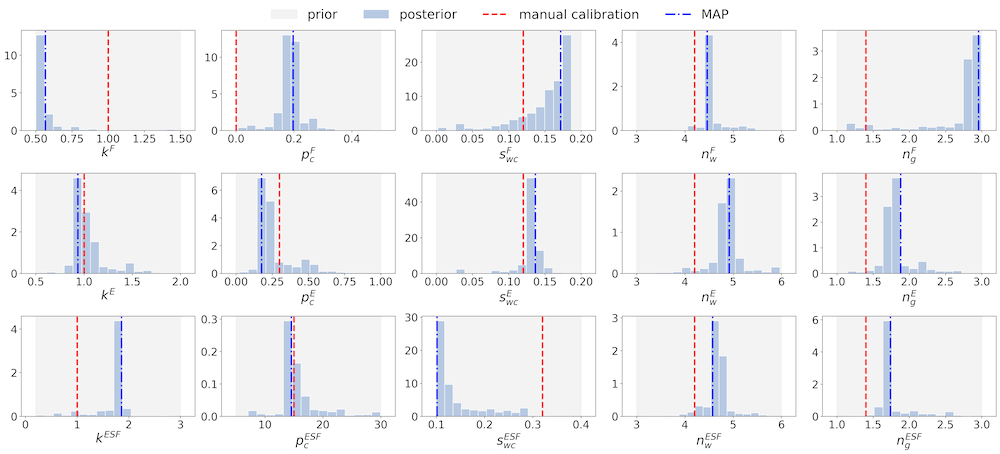}
\caption{Top: Comparison of experimental observations and MRST simulations evaluated at the inferred MAP parameters obtained from surrogate-assisted MCMC inversion. Bottom: Marginal posterior distributions obtained from MCMC sampling with random initialization (seed 1). Subplots are arranged consistently with Table~\ref{tab:params}, with rows representing geological units and columns representing parameter categories. Gray shaded regions indicate prior support, blue histograms denote posterior samples, red dashed lines indicate manually calibrated parameter values, and blue dash-dotted lines indicate MAP estimates.}
\label{fig:calibration1}
\end{figure*}

\begin{figure*}[!htbp]%
\centering
\includegraphics[width = \textwidth, trim=0 0 0 555, clip]{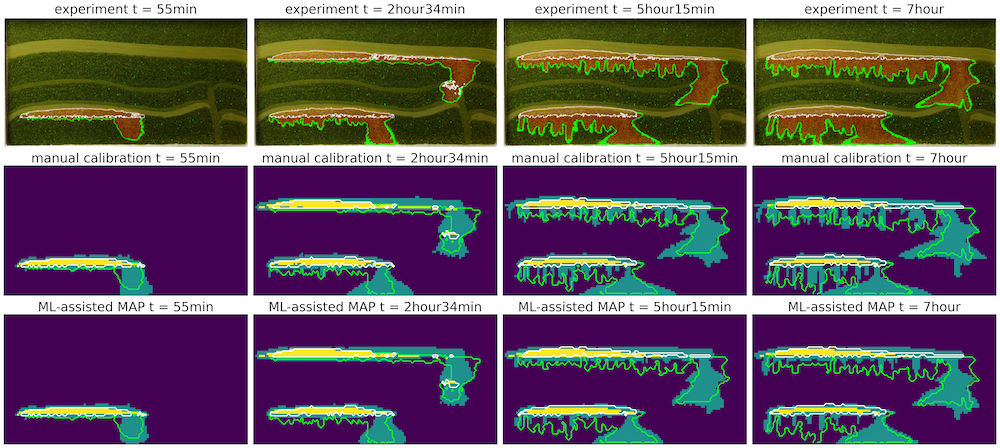}
\includegraphics[width = \textwidth]{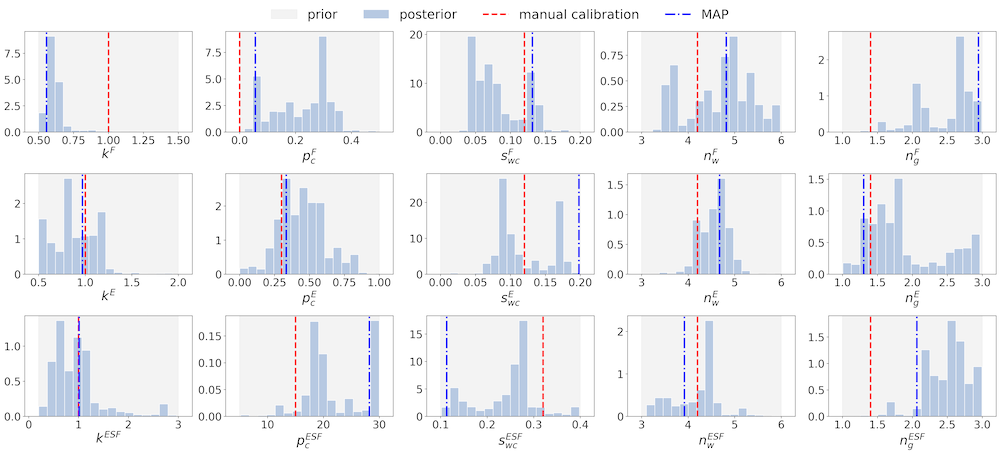}
\caption{Top: Comparison of experimental observations and MRST simulations evaluated at the inferred MAP parameters obtained from surrogate-assisted MCMC inversion. Bottom: Marginal posterior distributions obtained from MCMC sampling with random initialization (seed 2). Subplots are arranged consistently with Table~\ref{tab:params}, with rows representing geological units and columns representing parameter categories. Gray shaded regions indicate prior support, blue histograms denote posterior samples, red dashed lines indicate manually calibrated parameter values, and blue dash-dotted lines indicate MAP estimates.}
\label{fig:calibration2}
\end{figure*}

\end{appendices}

\end{document}